\documentclass[twocolumn,amsmath,amssymb,prl]{revtex4}
\usepackage{graphicx}% Include figure files
\usepackage{dcolumn}% Align table columns on decimal point
\usepackage{verbatim}
\usepackage{bm}
\newcommand{\Real}{\operatorname{{\mathrm Re}}}
\newcommand{\Imag}{\operatorname{{\mathrm Im}}}
\newcommand{\oiint}{\bigcirc \hspace{-0.5cm} \int \hspace{-0.25cm} \int}
\begin{document}
%\preprint{APS/123-QED}
%\usepackage{epsfig}
\title{Geometrical Rabi transitions between decoupled
  quantum states} \author{Xingxiang Zhou$^{1}$ and Ari
  Mizel$^{2}$} %\email{xizhou@yahoo.com}
\affiliation{ $^{1}$Laboratory of Quantum Information and Department
  of Physics, University of Science and Technology of China, Hefei,
  Anhui, 230026, China \\ 
$^2$SAIC, 4001 North Fairfax Drive, Suite 400, Arlington, VA 22203, USA
}
%\author{Xingxiang Zhou}
%\affiliation{ Laboratory of Quantum Information and Department
%  of Physics, University of Science and Technology of China, Hefei,
%  Anhui, 230026, China }
%\email{xizhou@yahoo.com}
 %  \author{Ari Mizel} %\email{ari@phys.psu.edu} 
%\affiliation{SAIC, 4001 North Fairfax Drive, Suite 400, Arlington, VA 22203, USA } 
\date{\today}

\begin{abstract}
  A periodic perturbation such as a laser field cannot induce
  transitions between two decoupled states for which the transition
  matrix element vanishes. We show, however, that if in addition some system
  parameters are varied adiabatically, such transitions become
  possible via adiabatic-change-induced excitations to other
  states. We demonstrate that full amplitude transfer between the two
  decoupled states can be achieved, and more significantly, the
  evolution of the system only depends on its path in parameter
  space. Our technique then provides a valuable means of studying
  nontrivial geometrical dynamics via auxiliary states with large
  energy splittings.

\end{abstract}


\maketitle

Central to the spirit of quantum mechanics are the concepts of
discrete quantum states and the transitions between them. To induce
coherent quantum transitions between two nondegenerate states, the
most often used technique is to apply an in-resonance periodic
perturbation such as a laser or microwave field.  This induces Rabi
oscillations \cite{ref:QO} if the transition matrix element of the
perturbation Hamiltonian between the two states is non-vanishing.
Rabi transitions are efficient in the sense that full amplitude
transfer can be achieved if the periodic perturbation is exactly in
resonance.  A different way of inducing quantum transitions is by
adiabatically varying some parameters in the system Hamiltonian
\cite{ref:Messiah}.  This makes the eigenstates of the system time
dependent and causes a system initially prepared in one eigenstate to
transition to other states.  Though inefficient in exciting the system
out of its initial state, adiabatic-change-induced quantum transitions
have some intriguing properties. In some important contexts such as
charge transfer via quantum pumping \cite{ref:Thouless83} in
solid-state physics, the system dynamics turn out to be geometrical,
depending only on the path traversed in parameter space and thus
allowing robust control of the system and precise transfer of charge.

\begin{figure}[h]
    \centering
    \includegraphics[width=3in, height=1.1in]{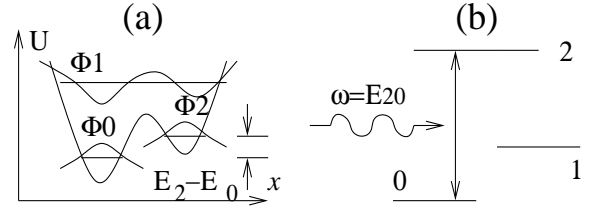}
    \caption{(a) Two localized and non-overlapping states $\Phi_0$ and
      $\Phi_2$ in a double-well potential. Here $x$ is the coordinate
      of the physical system under consideration (e.g., position for a
      quantum dot system and flux for a SQUID system). (b)
      Nondegenerate model system consisting of two decoupled states 0
      and 2 and auxiliary state 1. The periodic perturbation is in
      resonance with the 0-2 transition but off resonance with 0-1 and
      1-2 transitions.}

    \label{fig:double-well}
\end{figure}

In this work we consider the possibility of Rabi transitions between
two decoupled (and nondegenerate) quantum states, a problem of both
theoretical interest and practical significance. One such example is
the double-well potential depicted in Fig. \ref{fig:double-well} (a)
which has two localized states $\Phi_0(x)$ and $\Phi_2(x)$ (with
energies $E_0$ and $E_2$). Physical realizations of such a situation
can be a double-well quantum dot system \cite{ref:Qdot}, an SQUID
system biased close to half flux quantum \cite{ref:Squid}, or an
atomic system trapped in an optically engineered potential
\cite{ref:BEC}. If we apply a resonant periodic perturbation to a
system with two decoupled states, direct quantum transitions cannot
occur because the transition matrix element vanishes (which is the
definition of ``decoupled'').  In the double-well potential system in
Fig.  \ref{fig:double-well} (a), this is manifested by the fact that a
resonant perturbation $H'(x)\cos(E_2-E_0)t/\hbar$ has no effect since
$H'_{02}=0$ (which follows from the fact that $\Phi_0(x)$ and
$\Phi_2(x)$ do not overlap). Nevertheless, we explore the interesting
possibility of adiabatically changing some parameters of the system
Hamiltonian, in addition to applying a resonant periodic perturbation.
We will see that, by making use of inefficiently-excited auxiliary
states, full amplitude transfer between the two decoupled states can
be achieved. More significantly, the system dynamics is geometrical,
meaning it is dictated by rotation angles determined by the path the
system traversed in parameter space only and does not depend on time
explicitly.

We start by considering the simplest setup that consists of three
nondegenerate states as shown in Fig. 1 (b). These are assumed to be a
subspace of a system with unperturbed Hamiltonian $H_0(\bm{\lambda})$.
Here, $\bm{\lambda}$ is a collection of parameters. The eigenenergies
and eigenstates of $H_0(\bm{\lambda})$ are $E_n(\bm{\lambda})$ and
$\Phi_n(\bm{\lambda})$: $H_0(\bm{\lambda})\Phi_n(\bm{\lambda})=
E_n(\bm{\lambda})\Phi_n(\bm{\lambda})$, where $n$ is a set of quantum
numbers to label the spectrum of the system. We will use $n=0,1,2$ to
label our three state system. All other states are assumed to be
energetically far away from our three-state subspace and do not need
to be considered for our problem.

We apply to the system a (quasi) periodic perturbation $H'(t)=
2H'(\bm{\lambda})cos(\int \omega(t)dt)$ in resonance with state 0 and
2.  However, as discussed before, we assume that these two states are
decoupled, $\langle\Phi_0|H'|\Phi_2\rangle=0$.  The slow time varying
frequency $\omega(t)$ of the perturbation is close to the energy
difference between state 0 and 2. Such quasi periodic perturbation can
be realized, for instance, by applying a laser or microwave with a
time varying phase.  The Hamiltonian of the 3 state system is then
\begin{equation}
H=\sum_{j=0}^2 E_j(\bm{\lambda})|\Phi_j(\bm{\lambda}) \rangle 
\langle \Phi_j(\bm{\lambda}) | +2H'cos(\int
\omega(t)dt).
\label{eq:Eigen}
\end{equation}

Since state 0 and 2 are decoupled, the periodic perturbation, though
in resonance, cannot induce Rabi oscillations between them. To
facilitate possible transitions between state 0 and 2, we slowly vary
the parameters in the system Hamiltonian $H_0(\bm{\lambda})$, making
$\bm{\lambda}=\bm{\lambda}(t)$ time dependent.  When we use the
Schr\"odinger equation to solve for the wave function of the system,
$\psi(t)=\sum_{j=0}^2 c_j(t) \Phi_j(\bm{\lambda}(t))$, the time
dependence of the basis states must be taken into account.  Assuming
there is no Berry phase ($\langle\Phi_j (\bm{\lambda})|\dot{\Phi}_j
(\bm{\lambda}) \rangle= 0$, $\dot{\Phi}_j$ the time derivative of
${\Phi}_j$) \cite{ref:Thouless83}, we derive
%\begin{figure*}
%\begin{strip}
\begin{widetext}
\begin{equation}
i\hbar\frac{d}{dt}\left(
\begin{array}{c}
c_0 \\ 
c_1 \\
c_2
\end{array}
\right)=
\left(
\begin{array}{ccc}
  E_0(\bm{\lambda}) & H'_{01}(t)-i\hbar\langle\Phi_0|\dot{\Phi}_1\rangle & 0 \\
  H'_{10}(t)-i\hbar\langle\Phi_1|\dot{\Phi}_0\rangle & E_1(\bm{\lambda}) & H'_{12}(t) 
  -i\hbar\langle\Phi_1|\dot{\Phi}_2\rangle \\
  0 & H'_{21}(t)-i\hbar\langle\Phi_2|\dot{\Phi}_1\rangle & E_2(\bm{\lambda})
\end{array}
\right)
\left(
\begin{array}{c}
c_0 \\ 
c_1 \\
c_2
\end{array}
\right),
\label{Eq: 3-state}
\end{equation}
\end{widetext}
%\end{strip}
%\end{figure*}
where $H'_{1j} (\bm{\lambda}) =\langle\Phi_1 (\bm{\lambda}) |H'|\Phi_j
(\bm{\lambda})\rangle$, $j=0,2$ is the transition matrix element
between the auxiliary state $\Phi_1(\bm{\lambda})$ and the other two
states. Since $\hbar\omega$ is close to $E_2-E_0$ and off resonance
with $E_1-E_0$ and $E_2-E_1$, $H'_{1j}\neq 0$ does not imply direct
Rabi transitions between state $\Phi_1$ and $\Phi_j$. Notice the terms
$\hbar\langle\Phi_1 (\bm{\lambda})|\dot{\Phi}_j (\bm{\lambda}) \rangle$
also couple $\Phi_0(\bm{\lambda})$ and $\Phi_2(\bm{\lambda})$ to the
auxiliary state $\Phi_1(\bm{\lambda})$.

There are a few arguments that we can use to simplify the equations of
motion and understand the system dynamics. First, the precise
conditions for ``adiabatic'' change and ``off resonance'' transitions
are specified as $|\hbar\langle\Phi_j|\dot{\Phi}_1\rangle/(E_1-E_j)|$,
$|H_{1j}'/(|E_1-E_j|-\hbar\omega)|<<1$ for $j=0,2$.  Under such
conditions, the transitions from state 0 and 2 to state 1 due to
adiabatic change of the system Hamiltonian and off-resonance Rabi
oscillation are very inefficient. Consequently, the amplitude on state
1 remains small and follows those on state 0 and 2
adiabatically. Second, when we study the amplitudes on state 0 and 2,
it is convenient to switch to the ``rotating frame'' defined by $a_0=
c_0$, $a_2=\exp\{i\int \omega(t) t\}c_2$ (we have chosen $E_0$ as the
reference energy).  In doing so, we can use the rotating wave
approximation to drop fast oscillating terms in the rotating frame.
We are led to the following effective equations for the amplitudes on
state 0 and 2:
\begin{equation}
  i\hbar\frac{d\psi_R}{dt}
  =
  \left(
    \begin{array}{cc}
      \delta E_0(\bm{\lambda},t) & \hbar\kappa(\bm{\lambda},t) \\
      \hbar\kappa(\bm{\lambda},t)^* & \delta E_2(\bm{\lambda},t)+\Delta
    \end{array}
  \right)
  \psi_R.
  \label{eq:effective}
\end{equation}
Here, the two component wavefunction $\psi_R=(a_0,a_2)^{T}$, the Stark
shifts are given by $\delta E_j = \left[ 2|H'_{1j}|^2+
  |\hbar\langle\Phi_j| \dot{\Phi}_1 \rangle|^2 \right] /(E_j-E_1)$,
$j=0,2$, the detuning by $\Delta =E_2-E_0-\hbar\omega$, and the
effective Rabi frequency by
\begin{equation}
  \kappa= 
  \frac{i\langle\Phi_0|
    \dot{\Phi}_1 \rangle \langle \Phi_1| H'|\Phi_2 \rangle  }{E_1-E_0} 
  +\frac{i   \langle \Phi_0| H'|\Phi_1 \rangle \langle\Phi_1|
    \dot{\Phi}_2 \rangle}{E_1-E_2}.
\label{Eq: effect-Rabi-freq}
\end{equation}
When there are more than one auxiliary states, their contributions can
simply be summed.

If we choose the (time varying) frequency of the laser appropriately
so that $\Delta=\delta E_0(\bm{\lambda},t) -\delta
E_2(\bm{\lambda},t)$, the equal diagonal terms in Eq.
(\ref{eq:effective}) can be dropped because they give an unobservable
overall phase to $\psi_R$.  Using $\dot{\Phi}_j =\nabla_{\bm{\lambda}}
\Phi_j\cdot d\bm{\lambda} /dt$, $j=1,2$, we can rewrite
Eq. (\ref{eq:effective}) as follows:
\begin{equation}
id\psi_R=\{[\Real \bm{f}(\bm{\lambda})]\sigma_x - [\Imag \bm{f}(\bm{\lambda})] \sigma_y\}\psi_R\cdot d\bm{\lambda}.
\label{eq:geometrical}
\end{equation}
Here, $\Real \bm{f}$ and $\Imag \bm{f}$ are the real and
imaginary part of $\bm{f}(\bm{\lambda})$, a function of the parameters
$\bm{\lambda}$ only:
\begin{equation}
  \bm{f}
  =\frac{i \langle\Phi_0|
    \nabla_{\bm{\lambda}} \Phi_1 \rangle 
    H'_{12}}{E_1-E_0} 
  +\frac{i 
    H'_{01} \langle\Phi_1|
    \nabla_{\bm{\lambda}} \Phi_2 \rangle }{E_1-E_2}.
\label{eq:effect_rabi_freq}
\end{equation}

The intriguing dynamics described by Eqs. (\ref{eq:geometrical}) and
(\ref{eq:effect_rabi_freq}) are our main results. These equations
describe the rotation of an effective spin 1/2 under the effective
magnetic field determined by $\Real \bm{f}$ and $\Imag \bm{f}$.
However, remarkably, the evolution of the wavefunction is solely
determined by the path of $\bm{\lambda}$ in the parameter
space. Therefore, the system dynamics is geometrical (and non-Abelian
in general), even though the Berry phase has been set to zero
($\langle\Phi_j (\bm{\lambda})|\dot{\Phi}_j (\bm{\lambda}) \rangle=
0$) \cite{ref:nonAbelian}.

Formally, The solution of Eq. (\ref{eq:geometrical}) is $\psi_R(t) =P
\exp{\left( -i\int \{ [\Real \bm{f}(\bm{\lambda})]\sigma_x - [\Imag
    \bm{f}(\bm{\lambda})] \sigma_y\} \cdot d\bm{\lambda} \right)}
\psi_R(0)$, where $P$ stands for path ordering. When $\langle\Phi_0|
\nabla_{\bm{\lambda}} \Phi_1
\rangle %\langle \Phi_1| H'|\Phi_2 \rangle
H'_{12}$ and $H'_{01} \langle\Phi_1| \nabla_{\bm{\lambda}} \Phi_2
\rangle$ are real, Eq. (\ref{eq:geometrical}) can be integrated
explicitly.  The solution is $\psi_R(t)=\exp{\left(-i\Gamma \sigma_y
  \right)} \psi_R(0)$, where $\Gamma$ is a geometrical angle
determined by the path of the adiabatically varied parameters
$\bm{\lambda}(t)$ only: $\Gamma=i \int \bm{f}(\bm{\lambda}) \cdot
d\bm{\lambda}$. The physics is similar to conventional resonant Rabi
oscillation which can be understood in terms of an effective spin 1/2
rotating around the $y$ axis. However, rather than oscillating
sinusoidally with time, the amplitudes on the two states are
determined by a geometrical rotation angle not explicitly dependent on
time. When the parameters undergo cyclic changes, their paths form
closed loops in parameter space and the phase $\Gamma$ can be
expressed as a surface integral using Stokes's theorem.  For instance,
in the case of two adiabatically varied parameters, $\bm{\lambda} (t)=
(\mu (t), \nu (t))$, which form a closed loop $C$ in the $\mu - \nu$
plane, $\Gamma$ can be expressed as
\begin{align*}
  \oiint _C \left\{ \left( \frac{\partial}{\partial\mu} \frac{H_{12}'
        \langle \Phi_0 |}{E_1-E_0} \right) \left | \frac{\partial
        \Phi_1} {\partial \nu} \right \rangle - \left(
      \frac{\partial}{\partial\nu} \frac{H_{12}' \langle \Phi_0
        |}{E_1-E_0} \right) \left | \frac{\partial \Phi_1} {\partial
        \mu} \right \rangle \right\} \\ %\nonumber \\ %d\mu d\nu \\
  +\left\{ \left( \frac{\partial}{\partial\mu} \frac{H_{01}' \langle
        \Phi_1 |}{E_1-E_2} \right) \left | \frac{\partial \Phi_2}
      {\partial \nu} \right \rangle - \left(
      \frac{\partial}{\partial\nu} \frac{H_{01}' \langle \Phi_1
        |}{E_1-E_2} \right) \left | \frac{\partial \Phi_2} {\partial
        \mu} \right \rangle \right\} d\mu d\nu, \nonumber
\end{align*}
where $\bigcirc \hspace{-0.4cm} \int \hspace{-0.2cm} \int _C$ denotes
the integral over a surface whose boundary is the closed parameter
path $C$. When $\Gamma= \pi/2$, a system initially prepared in state 0
will evolve into state 2. Therefore, complete population transfer
between these 2 states can be achieved, as in resonant Rabi
transitions.

In the above, we have demonstrated that by using an auxiliary state
and changing the parameters of the system Hamiltonian adiabatically,
transitions between two decoupled quantum states can be
induced. Remarkably, the resulting physics has advantages of both
conventional Rabi oscillations and quantum pumping. It is possible to
fully transfer the population between the two states, thus realizing
efficient transitions. Moreover, the transition process is dictated by
geometrical phases determined by the path of the adiabatically varied
parameters only, and the speed with which the parameters are varied is
irrelevant as long as the adiabatic conditions are satisfied. Our
technique then provides a valuable means of studying nontrivial
geometrical dynamics, with the aid of auxiliary states that have large
energy splittings.

\begin{figure}[h]
  \includegraphics[width=3.2in, height=2in]{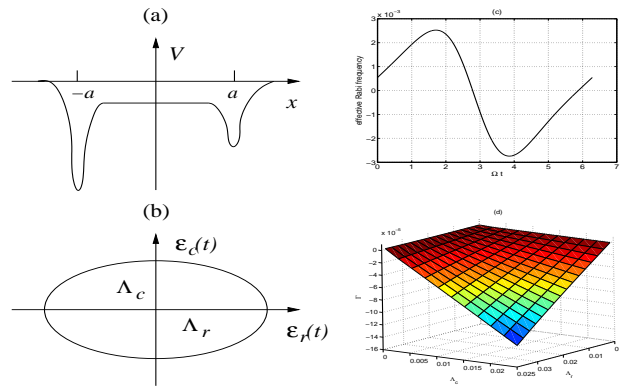}
  \caption{(a) Two delta-function potential wells with a square
    potential well in between.  (b) The elliptic and adiabatic path
    for the depths of the central and right potential wells:
    $\varepsilon_r(t)=\hbar^2\gamma_r(t)^2/2m=\hbar^2\gamma_r^2/2m+\Lambda_r
    \cos\Omega t$,
    $\varepsilon_c(t)=\hbar^2\beta(t)^2/2m=\hbar^2\beta^2/2m+\Lambda_c\sin\Omega
    t$. (c) The effective Rabi frequency in one cycle for
    $\Lambda_r=0.037E_u$, $\Lambda_c=0.024E_u$
    ($E_u=\hbar^2(\gamma_r^2-\beta^2)/2m$ used as the energy scale),
    in unit of $\mathcal{F}\zeta\Omega/E_u$. Only the few lowest bound
    states in the central well with a substantial contribution are
    included. (d) The rotation angle $\Gamma$ per cycle for different
    values of $\Lambda_r$ and $\Lambda_c$, in unit of
    $\mathcal{F}\zeta/E_u$.}
    \label{fig:delta_well}
\end{figure}

As a heuristic example, we consider a 1D problem of transferring a
particle between two localized delta-function potential wells via
extended states in a shallower square potential well. This is shown in
Fig. \ref{fig:delta_well} (a). The potential is given by
$V(x)=-V_c\theta(a-|x|) -U_l\delta(x+a) -U_r\delta(x-a)$, where
$\theta(x)$ is the unit step function, $2a$ is the distance between
the two delta-function potential wells, and $V_c$ and $U_{(l,r)}$ (all
positive) characterize the depths of the $c$entral, $l$eft and $r$ight
potential wells.  Such an idealized model with a small number of
parameters can be used to approximate many physical systems such as a
quantum dot or SQUID system with external biases to fine tune the
potential.

Assuming the left delta-function well is the deepest, we define a unit
length $\zeta=1/\gamma_l=\hbar^2/2mU_l$ ($m$ the ``effective mass'' of
the particle), the extension of the wavefunction of an isolated
delta-function potential well of depth $U_l$. When we choose
$a=44\zeta$, $\gamma_r=2mU_r/\hbar^2=22/a$,
$\beta=\sqrt{2mV_c}/\hbar=7.8/a$, the system has two localized bound
states in the two delta-function potential wells and a few unlocalized
bound states in the central well. Assuming the particle is initially
in the left well, we transfer it to the right well by applying a
periodic perturbation in resonance with the localized bound states and
adiabatically varying the depths of the central and right wells.

We assume a dipolar-like interaction between the applied field (of
strength $\mathcal{E}$) and the particle (with charge $q$).  The field
induced transition strength between states $i$, $j$ ($i,j$ = $l,c,r$)
is then $H'_{ij}=\mathcal{F}\cdot\langle x\rangle _{ij}$, where
$\mathcal{F}=q\mathcal{E}$ and $\langle x\rangle _{ij}$ is the
position matrix element evaluated between the two states. For the
parameters we chose, $H'_{lr}$ is negligibly small compared to
$H'_{lc}$ and $H'_{rc}$.  There is then no direct coupling between
states in the delta-function wells and the transition between them
occurs via coupling to the extended states in the central well.
Assuming that the depths of the central and right wells are varied
adiabatically and cyclically according to an elliptic path as depicted
in Fig. \ref{fig:delta_well} (b), we plot the effective Rabi frequency
$\kappa$ during a cycle in Fig. \ref{fig:delta_well} (c). Since
$\kappa$ is real, the system will evolve like an effective spin 1/2
rotating around a fixed axis.  $\kappa$ has a complicated time
dependence, in contrast to conventional Rabi oscillations with
constant Rabi frequencies. The total rotation angle per cycle, though,
does not depend on the frequency at which the depths are varied.
Plotted in Fig. \ref{fig:delta_well} (d) is the rotation angle per
cycle as a function of the size of the parameter path. Generally
speaking, the transition is more efficient when the parameters are
varied more in a cycle.

\begin{figure}[h]
    \centering
    \includegraphics[width=3in, height=1.4in]{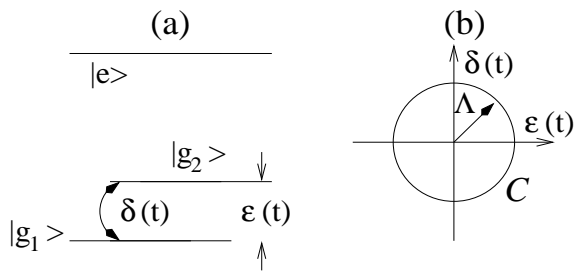}
    \caption{(a) A three-state system consisting of two coupled ground 
      states and an excited state. The energy difference $\varepsilon$
      and tunneling strength $\delta$ between the ground states are
      tunable. (b) A cyclic and circular path $C$ of $\varepsilon$ and
      $\delta$ in parameter space.}
    \label{fig:lambda_sys}
\end{figure}

Our new quantum transition mechanism is a general principle not
restricted to physical systems with localized states.  As another
example of its many possible applications, in the following we study a
problem in which quantum interference plays an essential role. We
consider a three-state system as shown in Fig.  \ref{fig:lambda_sys}
(a).  Here, the energies of the two ground states $|g_1\rangle$ and
$|g_2\rangle$ are both close to $E_g$. The energy splitting and
tunneling strength between them are $\varepsilon$ and $\delta$ which
are assumed to be tunable. An example of such a system is a Josephson
qubit \cite{ref:Squid} where $\varepsilon$ and $\delta$ are determined
by experimentally adjustable flux biases. The energy of an excited
state $|e\rangle$, $E_e$, is high above that of the ground states,
$E_e-E_g \gg \varepsilon, \delta$.  Therefore the unperturbed
Hamiltonian of the system is
\begin{equation}
H_0(\varepsilon,\delta)
=
\left(
\begin{array}{ccc}
E_g-\frac{\varepsilon}{2} & \frac{\delta}{2} & 0\\
\frac{\delta}{2} & E_g+\frac{\varepsilon}{2} & 0\\
0 & 0& E_e
\end{array}
\right).
\end{equation}
The upper-left block of the above Hamiltonian can be diagonalized to
obtain two eigenstates $|g_+\rangle =\sin\alpha |g_1\rangle
+\cos\alpha |g_2\rangle $, $|g_-\rangle =\cos\alpha |g_1\rangle
-\sin\alpha |g_2\rangle $ with eigenenergies $E_{\pm} =E_g\pm \sqrt{
  \varepsilon ^2+ \delta ^2}/2$, where $\alpha$ is defined by the
relations $\cos 2\alpha =\varepsilon /\sqrt{\varepsilon ^2+ \delta
  ^2}$ and $\sin 2\alpha =\delta /\sqrt{\varepsilon ^2+ \delta ^2}$.
If $\varepsilon$ and $\delta$ are varied adiabatically with time,
these eigenstates and eigenenergies become time dependent.

We assume that the system is prepared in the state $|g_-\rangle$ and a
periodic perturbation in resonance with $E_e-E_-$ is applied.  The
perturbative Hamiltonian is $H'(t)= \bm{d} \cdot 2\bm{\mathcal{E}}
\cos\int (E_e-E_-) dt/\hbar$, where $\bm{d}$ and $\bm{\mathcal{E}}$
are the analogues of the ``dipole moment'' operator and ``electric
field.''  We further assume that $\bm{d}_{eg_1}= \langle e | \bm{d}
|g_1 \rangle$ and $\bm{d}_{eg_2}= \langle e | \bm{d} |g_2 \rangle$ are
equal in magnitude and perpendicular in orientation: $|\bm{d}_{eg_1}|=
|\bm{d}_{eg_2}| =d$ and $\bm{d}_{eg_1} \perp \bm{d}_{eg_2}$
\cite{ref:Note}. If the angle between $\bm{d}_{eg_2}$ and the
polarization of $\bm{\mathcal{E}}$ is $\beta$, we find $H'_{eg_-}=
\langle e | \bm{d} |g_- \rangle \cdot \bm{\mathcal{E}} =d\mathcal{E}
\sin (\beta -\alpha)$ and $H'_{eg_+}= \langle e | \bm{d} |g_+ \rangle
\cdot \bm{\mathcal{E}}=d\mathcal{E} \cos (\beta -\alpha)$, where
$\mathcal{E}$ is the magnitude of the electric field.  When we choose
the polarization of the electric field such that $\beta= \alpha$,
$H'_{eg_-}= 0$ and $H'_{eg_+}=d\mathcal{E}$. In this case,
$|g_-\rangle$ and $|e \rangle$ are decoupled due to the destructive
interference. No transitions between these two states will occur even
though the frequency of the periodic perturbation is in resonance with
$E_e- E_-$.

In order to enable transitions between the two decoupled states $|g_-
\rangle$ and $|e \rangle$, we adiabatically vary $\varepsilon(t)$ and
$\delta(t)$.  This will induce (inefficient) transitions from $|g_-
\rangle$ to $|g_+ \rangle$ which is coupled to $|e \rangle$. The
condition for adiabaticity is $|\frac{\hbar\dot{\alpha}}{ \sqrt{
    \varepsilon ^2+ \delta ^2}}| \ll 1$.  Meanwhile, we adjust the
frequency of the periodic perturbation so that it remains in resonance
with $E_e- E_-$, and rotate the polarization of the $\bm{\mathcal{E}}$
field so that $\beta (t) =\alpha (t)$.  Though $|g_- (t)\rangle $ and
$|e \rangle$ remain decoupled due to the destructive interference,
transitions between them occur via adiabatic-change-induced excitation
to $|g_+ \rangle$.  If the system is initially in $|g_- \rangle$, and
$\varepsilon (t)$ and $\delta (t)$ undergo adiabatic and cyclic
changes along path $C$ in parameter space, the amplitude in the
excited state is $a_e = \sin{\Gamma}$, where $\Gamma$ is a geometrical
rotation angle given by
\begin{equation}
  \Gamma = 
  -d\mathcal{E}\int \frac{d\alpha}{\sqrt{ \varepsilon 
      ^2+ \delta ^2}} %\nonumber \\ 
  = \frac{d\mathcal{E}}{2} \oiint_C \frac{d\varepsilon d\delta}
  {(\varepsilon ^2+ \delta ^2)^{3/2}},
\end{equation}
where $\bigcirc \hspace{-0.4cm} \int \hspace{-0.2cm} \int _C$ denotes
the surface integral over the area bound by the closed path $C$.  As
an example, for a circular path of radius $\Lambda$ as depicted in
Fig. 2 (b), $\Gamma= -\frac{d\mathcal{E}}{\Lambda} \Delta \alpha $,
where $\Delta \alpha$ is the change of $\alpha$ along the path.

In conclusion, we have proposed a new mechanism for quantum
transitions which is realized by simultaneously applying a resonant
periodic perturbation and adiabatically changing the system
parameters. This new mechanism allows to realize quantum transitions
between decoupled states via inefficient excitations to auxiliary
states. Remarkably, it combines the advantages of previously known
methods, enabling efficient population transfer dictated by
geometrical angles. Aside from its fundamental interest, our scheme is
valuable for robust control of quantum transitions, and may find
applications in quantum information processing.

We thank K. Das for helpful discussions. This work was partly
supported by the Packard foundation. X. Z also acknowledges partial
support from National Natural Science Foundation of China (grant
No. 10875110).


\begin{references}
%\bibitem[$\dagger$]{byline}  Electronic address: xizhou@ece.rochester.edu.
%\bibitem[$\ddagger$]{ca}  Electronic address: zwzhou@ustc.edu.cn.

\bibitem{ref:QO} M. O. Scully and M. S. Zubairy, {\it Quantum
    Optics}, Cambridge University Press, 1997.  
\bibitem{ref:Messiah} A. Messiah, {\it Quantum Mechanics},  
    North-Holland, Amsterdam, 1962.
\bibitem{ref:Thouless83} D. J. Thouless, Phys. Rev. B {\bf 27}, 6083
  (1983).
\bibitem{ref:Qdot} Y. Masumoto and T. Takagahara, {\it Semiconductor
    Quantum Dots}, Springer, 2002.
%Phys. Rev. B {\bf 421}, 496 (1983). 
\bibitem{ref:Squid} T. P. Orlando {\it et al.},
%J. E. Mooij, L. Tian, Caspar H. van
%  der Wal, L. S. Levitov, S. Lloyd, and J. J. Mazo, 
  Phys. Rev. B {\bf 60}, 15398 (1999).  J. E. Mooij {\it et al.},
  Science {\bf 285}, 1036 (1999).
\bibitem{ref:BEC} S. Rolston, Phys. World {\bf 11}, 27 (1998).
\bibitem{ref:nonAbelian} M. V. Berry, Proc. R. Soc. Lond. A {\bf 392},
  45 (1984). F. Wilczek and A. Zee, Phys. Rev. Lett. {\bf 52}, 2111
  (1984).
\bibitem{ref:Note} We have made these assumptions to simplify our
  results.  The only condition required is that $\bm{d}_{eg_1}$ and
  $\bm{d}_{eg_2}$ have different orientations.


\end{references}
\end{document}